\documentclass[twoside,english,notitlepage]{iopart}
\usepackage[T1]{fontenc}
\usepackage[latin9]{inputenc}
\usepackage{geometry}
\usepackage{graphicx}

\makeatletter

\providecommand{\tabularnewline}{\\}

\usepackage{iopams}
\usepackage{setstack}

\usepackage[numbers, sort&compress]{natbib}

\usepackage{scrtime}

\makeatother

\usepackage{babel}

\begin{document}

\title{Finding critical points using improved scaling Ans\"{a}tze}

\author{M Roncaglia$^{1}$, L Campos Venuti$^{2}$, C Degli Esposti Boschi$^{3}$}

\address{$^{1}$ INRIM, Strada delle Cacce 91, 10135, Torino, Italy\\
\hspace{4pt} Politecnico di Torino, Corso Duca degli Abruzzi 24, 10129 Torino, Italy}

\address{$^{2}$ University of Southern California, Los Angeles, CA 90089, USA}

\address{$^{3}$ CNR-IMM, Sezione di Bologna, via Gobetti 101, 40129, Bologna, Italy}

\begin{abstract}
Analyzing in detail the first corrections to the scaling hypothesis,
we develop accelerated methods for the determination of critical points
from finite size data. The output of these procedures is sequences
of pseudo-critical points which rapidly converge towards the true
critical points. The convergence is faster than that obtained with
the fastest method available to date, which consists of estimating
the location of the gap's closure (the so called phenomenological
renormalization group). Having fast
converging sequences at our disposal, allows us to draw conclusions
on the basis of shorter system sizes. This can be extremely important
in particularly hard cases such as two-dimensional quantum systems
with frustrations or in Monte Carlo simulations when the sign problem
occurs. After reviewing the most efficient techniques available to
date, we test the effectiveness of the proposed methods both analytically
on the basis of the one-dimensional XY model, and numerically at phase
transitions occurring in non integrable spin models. In particular,
we show how a new Homogeneity Condition Method is able to produce fast converging sequences 
in correspondence of the Berezinskii-Kosterlitz-Thouless (BKT) transition simply making
use of ground-state quantities on relatively small systems. 
Remarkably, our method tested on the frustrated spin-$1/2$ Heisenberg model gives a BKT critical point 
which is incompatible with the ones present in the past literature, based on different methods. 
This discrepancy raises the fundamental question to determine the correct renormalization group approach    
and scaling assumptions that yield to the sequences converging to the true critical point. 
Finally, we formulate a general prescription that allows to analyze and 
efficiently locate critical points in a variety of cases, without knowing in advance
the universality class of the tested transition. 
Even if our methods are here tested in one dimension, we expect them to be valid in any spatial dimensionality
and both for quantum and classical statistical systems. 
\end{abstract}


\maketitle

\section{Introduction}

In statistical mechanics the exact evaluation of equilibrium properties,
except for non-interacting systems, can be worked out only in very
few notable integrable models \cite{baxter82}. Thanks to the increasing
power of computer resources, numerical simulations on finite size
systems are becoming a more and more popular. Among them we quote
exact diagonalization (ED) algorithms, density matrix renormalization
group (DMRG), and (quantum) Monte Carlo methods. The first question
one asks in this context is the determination of critical points in
order to classify the phase diagram into different phases. Once critical
points are found, one can ask what is the nature of the critical point
by computing critical exponents. Though statistical mechanics was
born more than a hundred years ago, and its theoretical understanding
is now on a sound basis after the scaling hypothesis and the theory
of renormalization group, until now only somehow phenomenological
methods are available for the determination of critical points from
finite size data. In this article we carefully analyze the leading
corrections of equilibrium averages beyond the usual scaling hypothesis.
In this way we are able to find accelerated methods for the determination
of critical points. The major advantage of having faster converging
methods is that one can content oneself with simulations on shorter
sizes. Such simulations are those which can be obtained with higher
accuracy in favorable cases, but are the only one available in the
most demanding situations given for instance by frustrated quantum
systems in two dimensions. 

Traditionally classical (some would say Euclidean) statistical systems,
have been studied by signaling out an order parameter (when present).
An example can be the magnetization in a magnetic system $\left\langle M\right\rangle $,
which, in the thermodynamic limit, is non zero only in the ordered
phase, and so allows to locate precisely the critical point. This
is only true in the thermodynamic limit, instead at finite sizes the
situation is less clear, and typically the magnetization will be different
from zero everywhere, presenting ''finite size tails'' in the paramagnetic
region. A possible, widely used, strategy is that of concentrating
on susceptibilities, i.e.~derivative of the magnetization. It can
happen that the susceptibility diverges at the critical point in the
infinite volume limit. At finite size singularity are smoothed and
the susceptibility only displays a maximum at a certain point which
is shifted respect to the actual critical point. Of course when one
considers larger and larger sizes, the pseudo-critical point given
by the maximum of the susceptibility, shifts towards the true critical
point. The convergence rate is algebraic in the system size $L$ and
is of the form $L^{-\theta}$ defining the so called shift exponent
$\theta$. This method of course is limited to those cases where a
susceptibility diverges. A somewhat more elaborated method is that
of the Binder's cumulant \cite{binder81}. In this approach one considers
the fourth order reduced cumulant $U_{L}\equiv1-\langle\left(\Delta M\right)^{4}\rangle/\left[3\langle\left(\Delta M\right)^{2}\rangle^{2}\right]$
where $\Delta M=M-\left\langle M\right\rangle $, in a finite system
of linear size $L$. Since by the scaling hypothesis $U_{L}$ should
depend only on the ratio $L/\xi$ where $\xi$ is the correlation
length, at the critical point $\xi\to\infty$ and $U_{L}$ should
be independent of $L$. The finite size pseudo critical point is then
obtained by solving $\partial_{L}U_{L}=0$ (here and in the following,
at finite size, derivatives with respect to $L$ must be interpreted
as finite differences). Graphically this amounts to finding the crossing
points of the curves $U_{L}$ at successive values of $L$. 

In quantum statistical systems the method of choice is given by the
so called phenomenological renormalization group (PRG) \cite{nightingale75,barber83}.
Continuous phase transitions in quantum mechanics are characterized
by closure of the first gap $\Delta_{L}$. Scaling hypothesis then
asserts that the gap must be a function of $\xi$ hence $\Delta_{L}\sim\xi^{-\zeta}$
which defines the dynamical exponent $\zeta$. Since at the critical
point the correlation length will be of the order of the system size,
the quantity $L^{\zeta}\Delta_{L}$ will be independent of $L$. The
PRG equation is then given by $\partial_{L}\left[L^{\zeta}\Delta_{L}\right]=0$. 

All these methods produce a sequence of pseudo critical points, say
$T_{L}^{\ast}$ which scale to the critical point according to an
algebraic law $T_{L}^{\ast}=T_{c}+O\left(L^{-\theta}\right)$ governed
by the shift exponent $\theta$. The general belief is that the shift
exponent $\theta$ is given by the inverse of the correlation length
exponent $\nu$, as it happens to the maximum of susceptibility. 
Although, to our knowledge, no systematic study of
shift exponents has been done, it was already pointed out in \cite{fisher82}
that the relation $\theta=1/\nu$ is not always valid, and $\theta$
depends, among other factors, on the boundary conditions. In particular
it should be clear that $\theta$ depends on the specific way in which
the pseudo-critical point has been obtained. 

It should be reminded here that in recent times new tools for the
characterization of quantum critical points, inherited from the theory
of quantum information, have arisen \cite{amico08}. Among these the
so called fidelity approach to phase transition consists of looking
for the maximum of the so called quantum Fisher information \cite{zanardi07,lcv07}.
This method is the best possible one if the only available resource
we admit is measuring observables. As we will see we can do much more,
i.e. compare different observables, compute finite difference and
so on. In any case it was shown in \cite{zanna08resource} that, except
for accidental cancellations, the shift exponent obtained with this
method is $\theta=1/\nu$. 

The lesson we should bear from this short resum\'e, is twofold. On the
one hand the present methods for the location of critical point are
somehow phenomenologically obtained assuming/imposing the scaling
hypothesis. On the other hand scaling hypothesis alone would imply
that finite size pseudo-critical points actually coincide with critical
points. To learn about the shift exponents and to develop faster converging
sequences one must go beyond the usual scaling hypothesis.

\section{Finite size scaling beyond scaling hypothesis\label{sec:scaling}}

To be concrete we will deal with quantum systems in $d$ spatial dimension
at zero temperature endowed with periodic boundary conditions (PBC)
to get rid of border effects. Our results can however be easily generalized
to finite temperature (in which case the role of the energy will be
played by the free energy) and/or classical statistical mechanics
in $D$ dimension (and use the dimensional crossover rule $D\leftrightarrow d+\zeta$).
The Hamiltonian can be written as 

\[
\mathcal{H}=\mathcal{H}_{0}+g\mathcal{V}=\mathcal{H}_{1}+\left(g-g_{c}\right)\mathcal{V}.\]

At the critical point $g_{c}$ the model is given by the Hamiltonian
$\mathcal{H}_{1}=\mathcal{H}_{0}+g_{c}\mathcal{V}$ which, by hypothesis,
is critical and so defines a scale invariant theory. The perturbation
$\mathcal{V}$ can then be expanded in terms of the scale-invariant
operators of the theory $\mathcal{H}_{1}$, this is the so called
hypothesis of local algebra \cite{patasinkij}. These operators are
classified into relevant/irrelevant in case their scaling dimensions
are smaller/larger than $d+\zeta$ , the case $d+\zeta$ being marginal.
Since, for $g\neq g_{c}$ the model is not critical, there must be
at least one relevant component in $\mathcal{V}$ which is responsible
for driving the system away from criticality. This simple observation
motivates the fact that, at least in principle, it should be possible
to estimate the critical point, only considering the finite size sequence
of the perturbation operator $b_{L}\equiv\left\langle \mathcal{V}\right\rangle /L^{d}$.
Such a strategy has been advocated in \cite{campos_local_06}, where
the so called finite size crossing method (FSCM) has been developed.
It was found that a sequence of pseudo-critical points $g_{L}^{\ast}$
could be given by the simple FSCM equation $\partial_{L}b_{L}\left(g_{L}^{\ast}\right)=0$.
As we will see later this method results in a shift exponent twice
as large as the ''standard'' one: $\theta_{FSCM}=2/\nu$. 

The energy density $e=\left\langle \mathcal{H}\right\rangle /L^{d}$
can be split into a regular and a singular part: $e=e_{\mathrm{reg}}+e_{\mathrm{sing}}$.
In the off-critical regime $L\gg\xi$ the singular part behaves as

\begin{equation}
e_{\mathrm{sing}}\sim\left|t\right|^{2-\alpha}\label{eq:e_sing}\end{equation}

where $t=g-g_{c}$. In the most general case, the coefficient in front
of Eq.~(\ref{eq:e_sing}) can depend on the side the transition is
approached, i.e.~$e_{\mathrm{sing}}\simeq A_{\pm}\left|t\right|^{2-\alpha}$
according to whether $t\to0^{\pm}$. Scaling hypothesis asserts that,
in a finite system of linear size $L$, thermodynamic averages depend
only on the ratio $L/\xi$, where, close to the transition point,
$\xi\sim\left|t\right|^{-\nu}$. Since for finite systems averages
must be analytic in $g$, the singular part of the energy must effectively
depend on the scaling variable $z=tL^{1/\nu}$. So we arrive at 

\begin{equation}
e_{\mathrm{sing}}=L^{-\left(d+\zeta\right)}\Phi_{0}\left(z\right)+\cdots,\label{eq:e_phi}\end{equation}
where the dots represent corrections arising from less singular contributions.
Here $\Phi_{0}\left(z\right)$ is a universal scaling function that
for $z\to\pm\infty$ (the off-critical regime) must behave as $\Phi_{0}\left(z\right)\sim A_{\pm}z^{\left(d+\zeta\right)\nu}$
in order to recover Eq.~(\ref{eq:e_sing}). Comparing with Eq.~(\ref{eq:e_sing})
one finds $2-\alpha=\left(d+\zeta\right)\nu$. Instead, for $L\ll\xi$
we are in the critical regime and $\Phi_{0}(z)$ behaves as an analytic
function. The only requirement is that, at finite $L$, $\Phi_{0}$
must be such that $e$ is an analytic function of $g$ also around
$g_{c}$. We will come back to this point later. 

Equation (\ref{eq:e_sing}) gives us the leading order singular part
of the free energy. Obviously one can expect corrections to such leading
behavior. One kind of corrections, so called \emph{analytic corrections,}
have the form:\begin{equation}
e_{\mathrm{sing}}\sim\left|t\right|^{2-\alpha}\left(1+a_{1}t+a_{2}t^{2}+\cdots\right).\label{eq:analytic}\end{equation}

Analytic corrections can be simply accounted for by noting that the
scaling variable $z$ is only approximately $z\approx tL^{1/\nu}$.
Using a more accurate prescription we can write $z=t\left(1+b_{1}t+b_{2}t^{2}+\ldots\right)L^{1/\nu}$.
Setting this ''improved'' scaling variable $z$ into Eq.~(\ref{eq:e_phi})
we obtain Eq.~(\ref{eq:analytic}) with $a_{1}=b_{1}\left(2-\alpha\right)$
and $a_{2}=\left(2-\alpha\right)\left(2b_{2}+b_{1}^{2}\left(1-\alpha\right)\right)/2$
\cite{aharony83}. This sort of analytic non-linearities are a consequence
of the fact that the RG equations are only approximately linear at
the fixed point. 

Besides analytic corrections, we can have corrections to Eq.~(\ref{eq:e_sing})
arising from irrelevant terms. This terms contribute to the singular
part of the free energy with an additive term of the form \cite{wegner72}\[
B_{\pm}h\left(t\right)\left|t\right|^{2-\alpha+\lambda},\]
with $\lambda$ not an integer and $h$ an analytic function of $t$
in zero. Such terms can be accounted for by a scaling function of
the form\[
L^{-\left(d+\zeta\right)-\omega}h\left(t\right)\Phi_{1}\left(z\right).\]
In the off-critical regime $z\to\pm\infty$, we must have $\Phi_{1}\left(z\right)\sim B_{\pm}z^{\left(d+\zeta+\omega\right)\nu}$
and we find $\lambda=\omega\nu$. Summarizing, the scaling form for
the singular part of the free energy, including only the least irrelevant
operator, is\begin{equation}
e_{\mathrm{sing}}=L^{-\left(d+\zeta\right)}\left[\Phi_{0}\left(z\right)+L^{-\omega}h\left(t\right)\Phi_{1}\left(z\right)\right].\label{eq:e_sing_FSS}\end{equation}

In the computation of the shift exponents, other terms in the expansion
of the energy density may play an important role. We write these terms
as\begin{equation}
F\left(g\right)L^{-\left(d+\zeta\right)}+\ldots.\label{eq:casimir-like}\end{equation}
Where the function $F$ is analytic in $g$ also around $g_{c}$.
Note that the next term in such an expansion would be of the form
$M\left(g\right)L^{-\left(d+\zeta+\epsilon\right)}$, with $M\left(g\right)$
also analytic. Such a term, close to the critical point is equivalent
to that arising from an irrelevant operator with $\omega=\epsilon$.
Therefore we do not include such terms as they are already included
in equation (\ref{eq:e_sing_FSS}).

In $d=1$, when the theory is conformally invariant (and so $\zeta=1$),
the origin of a $g$-dependent term of order $L^{-\left(d+\zeta\right)}$
like the one in (\ref{eq:casimir-like}), can be justified in the
following way. In this case it is known that \cite{affleck86}, at
the critical point, the energy density behaves as $e=e_{\infty}\left(g_{c}\right)-L^{-2}\pi cv/6+\cdots$.
Here $c$ is the central charge and $v$ the speed of elementary excitations.
Now, the effect of the perturbation, beside that of opening a gap,
will be that of renormalizing the speed of elementary excitations
$v\to v\left(g\right)$ giving rise to the function $F\left(g\right)=-\pi cv\left(g\right)/6$.
The presence of such terms can also be checked in the $XY$ model
via an explicit computation (see section \ref{sub:XY}). The appearance
of a $g$-dependent Casimir term, analogous to our $F\left(g\right)L^{-\left(d+\zeta\right)}$,
has also been demonstrated for the Casimir effect when adding a small
mass to the electromagnetic field \cite{barton84}.

Armed now with equation (\ref{eq:e_sing_FSS}) and possible corrections
of the form (\ref{eq:casimir-like}), we can now compute shift exponents
related to different techniques. The recipe is simple. What we have
to do is expand the equation determining the finite-size pseudo critical
point, in the quasi-critical region $L\ll\xi$. It suffice to expand
the equations up to linear terms in $\left(g-g_{c}\right)$. Higher
order terms in $\left(g-g_{c}\right)$ would result in corrections
which decay more rapidly with $L$ and so can be safely ignored since
we are interested in the leading behavior. 

An important issue in the computation of shift exponents is whether
the scaling function $\Phi_{0}$ (and/or $\Phi_{1}$) contain a linear
term or not. If a linear term is present in, say $\Phi_{0}$, it means
that the singular part of the energy will not be an even function
of $t$ close to the critical point. As a consequence, different curves
$e_{L}\left(g\right)$ at different sizes $L$ will cross close to
the critical point. In other words one could apply the FSCM equation
to the energy itself: $\partial_{L}e_{L}\left(g_{L}^{\ast}\right)=0$.
However this method gives rise to a poor convergence rate, with a
shift exponent given by $1/\nu$ as usual. The same shift exponent
is obtained if one applies the FSCM to the perturbation operator,
i.e.~solving $\partial_{L}b_{L}\left(g_{L}^{\ast}\right)=0$ as in
the original formulation \cite{campos_local_06}. 

Let us consider in detail the equation $\partial_{L}e_{L}\left(g_{L}^{\ast}\right)=0$.
Up to leading order the solution is given by\[
g_{L}^{\ast}=g_{c}+\frac{\left(d+\zeta\right)\nu\left(F\left(g_{c}\right)+\Phi_{0}\left(0\right)\right)}{\left(1-\left(d+\zeta\right)\nu\right)\Phi_{0}'\left(0\right)}L^{-1/\nu}.\]
We note that the leading convergence rate is dictated by the combination
$\left(F\left(g_{c}\right)+\Phi_{0}\left(0\right)\right)$ being non-zero.
In the energy density these terms correspond to quantities which scale
as $L^{-(d+\zeta)}$ in the quasi-critical region. We now present
a method which gets rid of such contributions and hence allows for
a faster convergence rate.

\subsection{The homogeneity condition method}

Since the convergence rate in the FSCM is dictated by $\left(F\left(g_{c}\right)+\Phi_{0}\left(0\right)\right)$
being non-zero, one can seek for a method such that the terms of order
$L^{-\left(d+\zeta\right)}$ cancels out exactly from the expression
of $e_{L}$. This is achieved by requiring that the function $\partial_{L}e_{L}$
be homogeneous of degree $\left(d+\zeta+1\right)$. We arrive in this
way at the \emph{homogeneity condition method} (HCM), which consists
of finding the solution to the following equation\begin{equation}
\left(d+\zeta+1\right)\partial_{L}e_{L}\left(g\right)+L\partial_{L}^{2}e_{L}\left(g\right)=0.\label{eq:HCM}\end{equation}
Taking into account the full Ansatz Eqs.~(\ref{eq:e_sing_FSS}) and
(\ref{eq:casimir-like}) we find that the convergence rate of the
pseudo-critical sequence so defined, at leading order is given by
\begin{equation}
g_{L}^{\ast,HCM}-g_{c}=\frac{\nu^{2}\omega\left(d+\zeta+\omega\right)\Phi_{1}\left(0\right)h\left(0\right)}{[\left(d+\zeta\right)\nu-1]\Phi_{0}'\left(0\right)}L^{-1/\nu-\omega}.\label{eq:HCM_shift}\end{equation}
We see that the shift exponent in this case is $\theta_{HCM}=\omega+1/\nu$.
It should be pointed out that, by considering higher derivatives of
$e_{L}$ with respect to $L$, the procedure could be iterated and
one can in principle find even faster converging sequences. However
then one would require explicit knowledge of other exponents which
are actually unknown. Note also that, despite the origin of the HCM
is that of eliminating a term of order $L^{-\left(d+\zeta\right)}$
from the expression of $e_{L}$, the HCM works irrespective of the
presence of such a term in $e_{L}$. 

The HCM equation (\ref{eq:HCM}) has been written using formally continuous
value of $L$. To obtain the HCM condition at finite size we first
replace derivatives with centered finite differences which have fast
convergence properties, so $D_{L}e_{L}\equiv\left(e_{L+\delta L}-e_{L-\delta L}\right)/\left(2\delta L\right)$
and $D_{L}^{2}e_{L}\equiv\left(e_{L+\delta L}-2e_{L}+e_{L-\delta L}\right)/\left(\delta L\right)^{2}$.
Then in principle any combination which reduces to (\ref{eq:HCM})
for $L\to\infty$ would produce a pseudo-critical sequence with shift
exponent given by $\theta_{HCM}$. However, at finite sizes, we can
require that the term of order $O\left(L^{-\left(d+\zeta\right)}\right)$
in (\ref{eq:casimir-like}) cancels out \emph{exactly}. This choice
will result in sequences with nicer converging properties. This is
achieved by choosing a proper combination of $D_{L}e_{L}$ and $D_{L}^{2}e_{L}$
, and the HCM equation at finite size becomes\begin{equation}
P_{1}\left(L\right)D_{L}e_{L}\left(g\right)+P_{2}\left(L\right)D_{L}^{2}e_{L}\left(g\right)=0\label{eq:HCML}\end{equation}
 where the two polynomials $P_{1/2}$ are given by\begin{eqnarray*}
P_{1}\left(L\right) & = & \left[2L^{d+\zeta}\left(L+\delta L\right)^{d+\zeta}+2L^{d+\zeta}\left(L-\delta L\right)^{d+\zeta}-4\left(L^{2}-\delta L^{2}\right)^{d+\zeta}\right]\\
P_{2}\left(L\right) & = & \delta L\left[L^{d+\zeta}\left(L+\delta L\right)^{d+\zeta}-L^{d+\zeta}\left(L-\delta L\right)^{d+\zeta}\right].\end{eqnarray*}
For instance, when $d+\zeta=2$ the HCM equation reduces to $\left[3L^{2}-\left(\delta L\right)^{2}\right]D_{L}e_{L}\left(g\right)+L^{3}D_{L}^{2}e_{L}\left(g\right)=0$,
which correctly reduces to (\ref{eq:HCM}) in the limit of large $L$.

\subsubsection{No linear terms in the scaling functions.}

As can be inferred by looking at equation (\ref{eq:HCM_shift}), the
HCM \emph{as is} cannot be applied when the scaling function $\Phi_{0}$
does not contain a linear term. In this case one would have $\Phi_{0}'\left(0\right)=0$
making equation (\ref{eq:HCM_shift}) meaningless. However in this
situations, finite size corrections are smaller, and so it should
be possible to envisage a method to turn this point into advantage.
This is easily achieved by applying the HCM equation (\ref{eq:HCM})
to the perturbation $b_{L}$ rather than to the energy density. In
this case the solution of $\left(d+\zeta+1\right)\partial_{L}b_{L}\left(g\right)+L\partial_{L}^{2}b_{L}\left(g\right)=0$
gives\[
g_{L}^{\ast,HCM}-g_{c}=\frac{\left(1-\nu\omega\right)\left(1-\nu\left(d+\zeta+\omega\right)\right)h\left(0\right)\Phi_{1}'\left(0\right)}{[2(d+\zeta)\nu-4]\Phi_{0}''\left(0\right)}L^{-1/\nu-\omega},\]
and the shift exponent is $1/\nu+\omega$. If also $\Phi_{1}$ does
not contain a linear term the shift exponents becomes $2/\nu+\omega$
when no other cancellations are present. The numerical test we have
performed (see section \ref{sub:Numerical-example} below) fall in
this case (see also table \ref{tab:summary}). For completeness we
note that, when $\Phi_{0}'\left(0\right)=0$, the FSCM applied to
the energy gives formally a shift exponent $1/\nu-\omega$. This is
clearly a poor convergence and the method can be successfully applied
only when $1/\nu>\omega$.

\subsection{The Berezinskii-Kosterlitz-Thouless transition}

The presence of the additional $\omega$ in $\theta_{HCM}$ suggests
that the method could work even in the situation where formally $\nu=\infty$
like in the Berezinskii-Kosterlitz-Thouless (BKT) transition. In the
BKT case the correlation length diverges with an essential singularity
of the form $\xi\sim\exp\left(at^{-\sigma}\right)$, where $a$ and
$\sigma$ are constants. For instance in the classical two dimensional
XY model $\sigma=1/2$. Using a modified Ansatz adapted to the BKT,
one can show \cite{roncaglia08} that the HCM Eq\@.~(\ref{eq:HCM})
still defines a sequence of pseudo-critical points converging to the
true critical point with shift exponent $\theta_{HCM}^{BKT}=\epsilon/\left(n-1+\sigma\right)$
with $\epsilon>0$ and $n$ is an integer greater than one. 

For the application of the HCM equation one simply requires the knowledge
of $b_{L}$ which is the average of a (typically local) quantity over
the ground state, and the dynamic exponent $\zeta$. We point out
here that, contrary to PRG method (the current method of choice for
estimating quantum critical points) no need of excited states is required.
This is a big advantage since, in every numerical method excited states
(if available) are typically obtained with smaller accuracy than the
ground state. Instead the local average $b_{L}$ can be obtained with
very high precision. In addition, the HCM is not restricted to quantum
mechanical systems and is superior to the PRG in that it produces
a faster converging sequence.

\begin{table}
\noindent \begin{centering}
\begin{tabular}{|c|c|c|c|}
\hline 
Method & $\Phi_{0}$ linear & $\Phi'_{0}(0)=0$ & $\Phi'_{0}(0)=0,\Phi'_{1}(0)=0$\tabularnewline
\hline
\hline 
FSCM($e_{L}$) & $1/\nu$ & $1/\nu-\omega$ or not applicable & not applicable\tabularnewline
\hline 
FSCM($b_{L}$) & $1/\nu$ & $\min\left(2/\nu,1/\nu+\omega\right)$ & $2/\nu$\tabularnewline
\hline 
HCM($e_{L}$) & $1/\nu+\omega$ & $1/\nu$ & not applicable\tabularnewline
\hline 
HCM($b_{L}$) & $1/\nu$ & $1/\nu+\omega$ & $2/\nu+\omega$\tabularnewline
\hline 
MHC & $1/\nu$ & $1/\nu+\omega$ & $2/\nu+\omega$\tabularnewline
\hline
\end{tabular}
\par\end{centering}

\caption{Summary of values of shift exponents in different cases. Only the
minimum possible shift exponent is given. For instance, when also
$\Phi_{1}'\left(0\right)=0$, the shift exponent for HCM($b_{L}$)
(which means HCM applied to $b_{L}$) and MHC becomes at least $2/\nu+\omega$.
When the method formally returns a sequence that does not converge
to the critical point in the limit $L\to\infty$ we have reported
{}``not applicable''. \label{tab:summary}}

\end{table}

\subsection{The modified homogeneity conditions}

The dynamic exponent $\zeta$ is one in most known cases, however
when not known \textit{a priori} it can be easily found by looking at the gap behavior
$\Delta=\mathrm{const.}\times L^{-\zeta}$ via exact diagonalization
on small lattices. However in some cases (like two-dimensional quantum
mechanics) assessing $\zeta$ can be a problem, but this requirement
can in fact be lifted. Considering also the finite size sequence of
the ground state energy $e_{L}$, it is possible to find a modified
homogeneity condition (MHC) given by\begin{equation}
\frac{b_{L}(g_{L}^{*})-b_{L-\delta L}(g_{L}^{*})}{e_{L}(g_{L}^{*})-e_{L-\delta L}(g_{L}^{*})}=\frac{b_{L+\delta L}(g_{L}^{*})-b_{L}(g_{L}^{*})}{e_{L+\delta L}(g_{L}^{*})-e_{L}(g_{L}^{*})},\label{eq:MHC}\end{equation}
or in the continuum version \begin{equation}
\frac{\partial_{L}b_{L}(g)}{\partial_{L}e_{L}(g)}=\frac{\partial_{L}^{2}b_{L}(g)}{\partial_{L}^{2}e_{L}(g)}.\label{eq:crossing_fast}\end{equation}
Using the appropriate scaling Ansatz for the energy $e_{L}$ and the
perturbation $b_{L}$ one can show, that the MCH Eq\@.~(\ref{eq:MHC})
produces a pseudo-critical sequence $g_{L}^{\ast,MHC}=g_{c}+\mathrm{const.}\times L^{-\theta_{MHC}}$.
If $\Phi_{0}$ has a linear term, the shift exponent is only $1/\nu$,
growing to $1/\nu+\omega$ when $\Phi_{0}'\left(0\right)=0$, which
becomes $2/\nu+\omega$ if also $\Phi_{1}'\left(0\right)=0$.

\subsection{General prescription}

In table \ref{tab:summary} we summarize the predicted shift exponents
associated to different methods. Notice that only the minimal shift
exponents are computed. When one has additional cancellations in the
scaling functions, shift exponents tend to be higher. For example,
when neither $\Phi_{0}$ or $\Phi_{1}$ contain a linear term, the
shift exponent computed using the HCM and the MHC is $2/\nu+\omega$.
From the available information about the coefficients $\Phi_{0}'(0)$
and $\Phi_{1}'(0)$, possibly gathered by means of analytical arguments
(symmetries, dualities, etc.) and/or graphical inspection of the numerical
data, one should be able to use the best possible {}``recipe'' to
locate the critical point with the largest shift exponent. For instance,
if $\Phi_{0}$ carries a linear term in the scaling variable, then
the best strategy turns out to be the HCM applied to the energy density.

However, sticking to the HCM, one could also adopt a more {}``automatic''
procedure, in the sense that both $e_{L}$ and $b_{L}$ are computed
at once without any previous knowledge about the linear terms in the
scaling functions. Then, one of the two following situations will
occur: (i) Both HCM($e_{L}$) and $ $HCM($b_{L}$) converge to the
same critical point (within numerical uncertainties), meaning that
$\Phi_{1}'(0)\neq0$ and one has to refer to the second or third column
of table \ref{tab:summary}. A best fit on the sequence with faster
convergence then yields the exponent $1/\nu+\omega$, while the other
sequence gives $1/\nu$ and so the two exponents $\nu$ and $\omega$
can be estimated independently. (ii) If instead HCM($e_{L}$) and
$ $HCM($b_{L}$) converge to different points (or the energy one
does not converge at all), then one has to rely on the latter and
the best fit returns $2/\nu+\omega$ (fifth column of table \ref{tab:summary}).
Even if now $\nu$ and $\omega$ are no longer separated, the convergence
towards the critical point is enhanced by a larger exponent.

This summary concludes our theoretical analysis of the improved shift
exponents based on the contributions beyond the scaling hypothesis.
In the following we will consider practical implementations of the
accelerated methods HCM and MHC and compare them against previous
methods namely the PRG and the FSCM. We will consider an exactly solvable
case where all the quantities can be calculated analytically, a non-integrable
example and finally one instance of the extremely severe BKT case.
In all cases the HCM and MHC represent an improvement with respect
to previous methods and allow one to reduce to shorter sizes. 

Our general strategy is that of obtaining accurate numerical values possibly free from uncontrollable errors, for short sizes. 
This can be achieved with Lanczos diagonalization on short sizes or DMRG keeping a large number of states and moderate sizes. 
In this way we avoid the pitfall of relying the extrapolated critical points on data 
which have potentially unknown errors as it may happen in quantum Monte Carlo
or the so-called ``infinite size'' DMRG.

\section{Examples}

\subsection{An exactly solvable case: the $XY$ model in transverse field\label{sub:XY}}

To check our methods analytically we consider the one-dimensional
$XY$ model in transverse field \cite{lsm61}. It is given by the
following Hamiltonian

\begin{equation}
H=-\sum_{j=1}^{L}\left[\frac{(1+\eta)}{2}\sigma_{j}^{x}\sigma_{j+1}^{x}+\frac{(1-\eta)}{2}\sigma_{j}^{y}\sigma_{j+1}^{y}+h\sigma_{j}^{z}\right].\label{eq:HXY}
\end{equation}
 As usual we will employ PBC to avoid the appearance of surface corrections,
and we will use $L$ even. The model Eq.~(\ref{eq:HXY}) can be exactly
solved by first passing to spinless fermion via Jordan-Wigner transformation
and then diagonalizing the quadratic Hamiltonian with a suitable Bogoliubov
transformation. It turns out that, for all $\eta\neq0$, there is
a critical point at $h=h_{c}=1$ which belongs to the same universality
class as the two-dimensional classical Ising model. All quantities
of interest like finite size energy density $e_{L}$ or perturbation
$b_{L}$, and even lowest gap $\Delta_{L}$ can be computed exactly.
In the quasi-critical regime $\xi\gg L$ one is able to find asymptotic
expressions which are Laurent expansion in $L$. More details about
the calculations can be found in \cite{roncaglia08}, here we only
quote the results. The energy density at finite size in the quasi-critical
region has the following form
\begin{eqnarray}
e_{L}\left(h,\eta\right) & = & e_{\infty}\left(1,\eta\right)-\frac{\pi\left|\eta\right|}{6}L^{-2}-\frac{7\pi^{3}}{360}\frac{\left(4\eta^{2}-3\right)}{4\left|\eta\right|}L^{-4}\nonumber \\
 & + & \left(h-1\right)\left[b_{\infty}\left(1,\eta\right)-\frac{\pi}{12\left|\eta\right|}L^{-2}-\frac{7\pi^{3}(3-2\gamma^{2})}{2880\left|\eta\right|^{3}}L^{-4}\right]\nonumber \\
 & - & \frac{\left(h-1\right)^{2}}{2}\frac{\ln\left(L\right)+\gamma_{C}+\ln\left(8\left|\eta\right|/\pi\right)-1}{\pi\left|\eta\right|}+O\left(\left(h-1\right)^{3}\right).\label{eq:e_XY}
\end{eqnarray}
The Ising model has $\alpha=0$ and a logarithmic singularity which
at this level shows up in the $\ln L$ term at second order. The perturbation
$b_{L}$ has a similar expansion in the quasi-critical region:
\begin{eqnarray}
b_{L}\left(h,\eta\right) & =b_{\infty}\left(1,\eta\right)-\frac{\pi}{12\left|\eta\right|}L^{-2}-\frac{7\pi^{3}(3-2\gamma^{2})}{2880\left|\eta\right|^{3}}L^{-4}\nonumber \\
 & -\left(h-1\right)\frac{\ln\left(L\right)+\gamma_{C}+\ln\left(8\left|\eta\right|/\pi\right)-1}{\pi\left|\eta\right|}+O\left(z^{2}\right).\label{eq:b_XY}
\end{eqnarray}
 The logarithmic singularity in principle requires a modified scaling
Ansatz than that outlined in section \ref{sec:scaling} (see e.g.~\cite{roncaglia08,barber83}).
However, stopping at first order, we can still compare this expression
with our scaling Ansätze Eqns.~(\ref{eq:e_sing_FSS}) and (\ref{eq:casimir-like})
expanded in the quasi-critical region. We observe that, in this case
$\Phi_{0}'\left(0\right)=\Phi_{1}'\left(0\right)=0$. We can also
read off the function $F\left(h\right)$ (here $h$ plays the role
of $g$). Clearly $F\left(h\right)=-\pi\left|\eta\right|/6-\pi\left(h-1\right)/\left(12\left|\eta\right|\right)+O\left(h-1\right)^{2}$.
This is in perfect agreement with the conformal field theory (CFT)
formula $F\left(h\right)=-\pi cv\left(h\right)/6$ taking into account
that the central charge is $c=1/2$ and the one particle dispersion
is $\Lambda_{k}=2\sqrt{\eta^{2}+h^{2}+\left(1-\eta^{2}\right)\cos\left(k\right)^{2}+2h\cos\left(k\right)}$
giving rise to an $h$-dependent velocity $v\left(h\right)=2\left|\eta\right|+\left(h-1\right)/\left|\eta\right|+O\left(h-1\right)^{2}$
close to the critical point.

For what regards the terms of order $L^{-4}$, since $d+\zeta=2$
one deduces that $\omega=2$ (this exponent was also noted in \cite{salas00}).
Since $\nu=1$ one would conclude that this term arises from an irrelevant
operator of dimension $\left(2+2\right)=4$. A deeper understanding
of the finite size corrections of the $XY$ model at the critical
point ($h=1$) can be obtained from the point of view of perturbed
CFT. It was shown in \cite{reinicke87} that the leading finite-size
corrections of the critical $XY$ model with PBC have scaling dimension
3 (this is an operator belonging to the tower of the energy density
in CFT language) and 4 (tower of the identity). The operator with
scaling dimension 3 would contribute to the energy density at finite
size with a term of order $L^{-3}$. However at first order in perturbation
theory the relevant matrix element vanishes identically, the first
non-zero contribution is at second order giving rise to a $L^{-4}$
correction \cite{reinicke87}. Summarizing, the leading corrections
arising from irrelevant operators with scaling dimension 3 and 4 are
both of order $L^{-4}$ in accordance with Eq. (\ref{eq:e_XY}) for
$h=1$.

After this short digression we can use Eqns.~(\ref{eq:e_XY}) and
(\ref{eq:b_XY}) to calculate shift exponents associated to different
techniques. Let us consider the HCM first. By direct inspection one
sees that in this case $\Phi_{0}'\left(0\right)=\Phi_{1}'\left(0\right)=0$.
According to table \ref{tab:summary} the HCM conditions are applicable with the
perturbation $b_{L}$. Applying the HCM equations to Eq.~(\ref{eq:b_XY})
one obtains
\[
h_{L}^{\ast,HCM}=1+\frac{7\pi^{4}\left(2\eta^{2}-3\right)}{720\eta^{2}}L^{-4}+\ldots,
\]
 telling us that $\theta_{HCM}=4$. Since, from the exact solution
one has $\nu=1$ and we just recovered $\omega=2$, this result is
consistent with $\theta_{HCM}=\omega+2/\nu$ valid when $\Phi_{0}'\left(0\right)=\Phi_{1}'\left(0\right)=0$.
The modified method MHC gives instead 
\[
h_{L}^{*,MHC}=1-\frac{7\pi^{4}(2\eta^{4}+\eta^{2}-3)}{720\eta^{2}}L^{-4}+\ldots,
\]
 which again confirms $\theta_{MHC}=\theta_{HCM}=4$ for this case.
In this model we can also find exactly the shift exponent related
to the PRG method. All we need is the expression for the lowest gap
at finite size, close to the critical point which we recall here from
\cite{roncaglia08}
\begin{eqnarray}
\Delta_{L} & = & \left(h-1\right)+\frac{\pi\left(2\eta^{2}+h-1\right)}{4\left|\eta\right|}L^{-1}\label{eq:XY_gap}\\
 & + & \frac{\pi^{3}\left[3\left(h-1\right)-2\left(2+h\right)\eta^{2}+8\eta^{4}\right]}{192\left|\eta\right|^{3}}L^{-3}+O\left(L^{-5}\right).\nonumber 
\end{eqnarray}
 Since $d=\zeta=1$ the PRG method consists of imposing $\partial_{L}\left[L\Delta_{L}\left(h_{L}^{\ast}\right)\right]=0$,
which yields \cite{roncaglia08}
\[
h_{L}^{*,PRG}=1+\frac{\pi^{3}\left(4\eta^{2}-3\right)}{48\left|\eta\right|}L^{-3}+\ldots.
\]

Finally the FSCM gives $h_{L}^{*}=1+L^{-2}\pi^{2}/6$, in accordance
with $\theta_{FSCM}=2/\nu$.

All in all these explicit calculations confirm that the accelerated
HCM and MHC methods are superior to the PRG or the FSCM methods.

\begin{figure}
\begin{centering}
\includegraphics[clip,width=10cm,height=6cm]{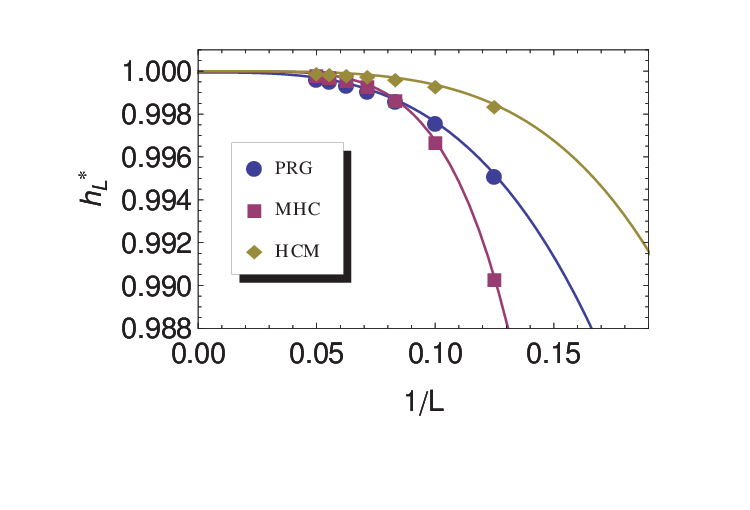} 
\par\end{centering}
\caption{(color on-line) Sequences of pseudo-critical points obtained with
the PRG, the MHC {[}given by Eq.~(\ref{eq:MHC}){]} and HCM {[}Eq.~(\ref{eq:HCML}){]}.
Continuous lines are algebraic best-fits to the data of the form $h_{L}^{*}=h_{c}+AL^{-\phi}$.\label{fig:XYfits}}
\end{figure}

In this exactly solvable case we can also see test the efficiency
of the various methods using finite size data on small sizes. Using
finite size data for $L=8,10,12,\ldots,20$ we compute sequences of
pseudo-critical points for the PRG, MHC and HCM methods. We then estimate
the critical point using a non-linear fit of the form $h_{L}^{*}=h_{c}+AL^{-\phi}$,
with $h_{c},\, A,\,\phi$ unknown. The results of the fit are plotted
in Fig.~\ref{fig:XYfits} while the estimated critical point and
errors are summarized in Table \ref{tab:XYfit}. 

\begin{table}
\noindent \begin{centering}
\begin{tabular}{|c|c|c|c|}
\hline 
Method  & Estimated $h_{c}$ & Error  & $\phi$\tabularnewline
\hline 
\hline 
PRG & $0.99994(2)$  & $9\times10^{-6}$ & $3.24$\tabularnewline
\hline 
HCM($b_{L}$)  & $1$  & $7\times10^{-8}$ & $4.03$\tabularnewline
\hline 
MHC  & $1$  & $2\times10^{-5}$  & $4.88$\tabularnewline
\hline 
\end{tabular}
\par\end{centering}

\caption{Summary of estimated critical points and error for the $XY$ model
in transverse field. The anisotropy has been fixed to $\eta=0.5$.
Pseudo-critical sequences are obtained using finite size data from
$L=8,10,12,\ldots,20$. The critical point is obtained using a non-linear
fit of the form $h_{L}^{*}=h_{c}+AL^{-\phi}$. The standard error
of the regression is defined according to $\left((N-p)^{-1}\sum_{j=1}^{N}(y_{j}-f(x_{j}))^{2}\right)^{1/2}$
where $(x_{j},y_{j})$ are the data points, the fitted function is
$f(x)=h_{c}+Ax^{\phi}$ and the number of parameters of the fit $p$
is 3 in our case. \label{tab:XYfit}}
\end{table}

\subsection{Numerical example: $c=1$ transition\label{sub:Numerical-example}}

We consider here a spin-1 model with anisotropies given by the following
Hamiltonian 
\begin{equation}
H=\sum_{j=1}^{L}\left[S_{j}^{x}S_{j+1}^{x}+S_{j}^{y}S_{j+1}^{y}+J_{z}S_{j}^{z}S_{j+1}^{z}+D(S_{j}^{z})^{2}\right],\label{eq:HJzD}
\end{equation}
 where as usual, PBC are used. The Hamiltonian Eq.~(\ref{eq:HJzD})
has been used to describe magnetic properties of different quasi 1D
compounds and an extensive literature is available (see e.g.~Refs.~\cite{cristian03,chen03,cristian04}
and reference therein). We fix here to the value $J_{z}=0.5$ where
a critical point of order $D_{c}=0.6\div0.7$ has been estimated previously.
At this particular transition point it has been previously found \cite{cristian03}
a pretty large value of the correlation length exponent $\nu=2.38$,
which implies a slow scaling in $1/\nu$.

\begin{figure}
\begin{centering}
\includegraphics[width=7cm,height=6cm,clip]{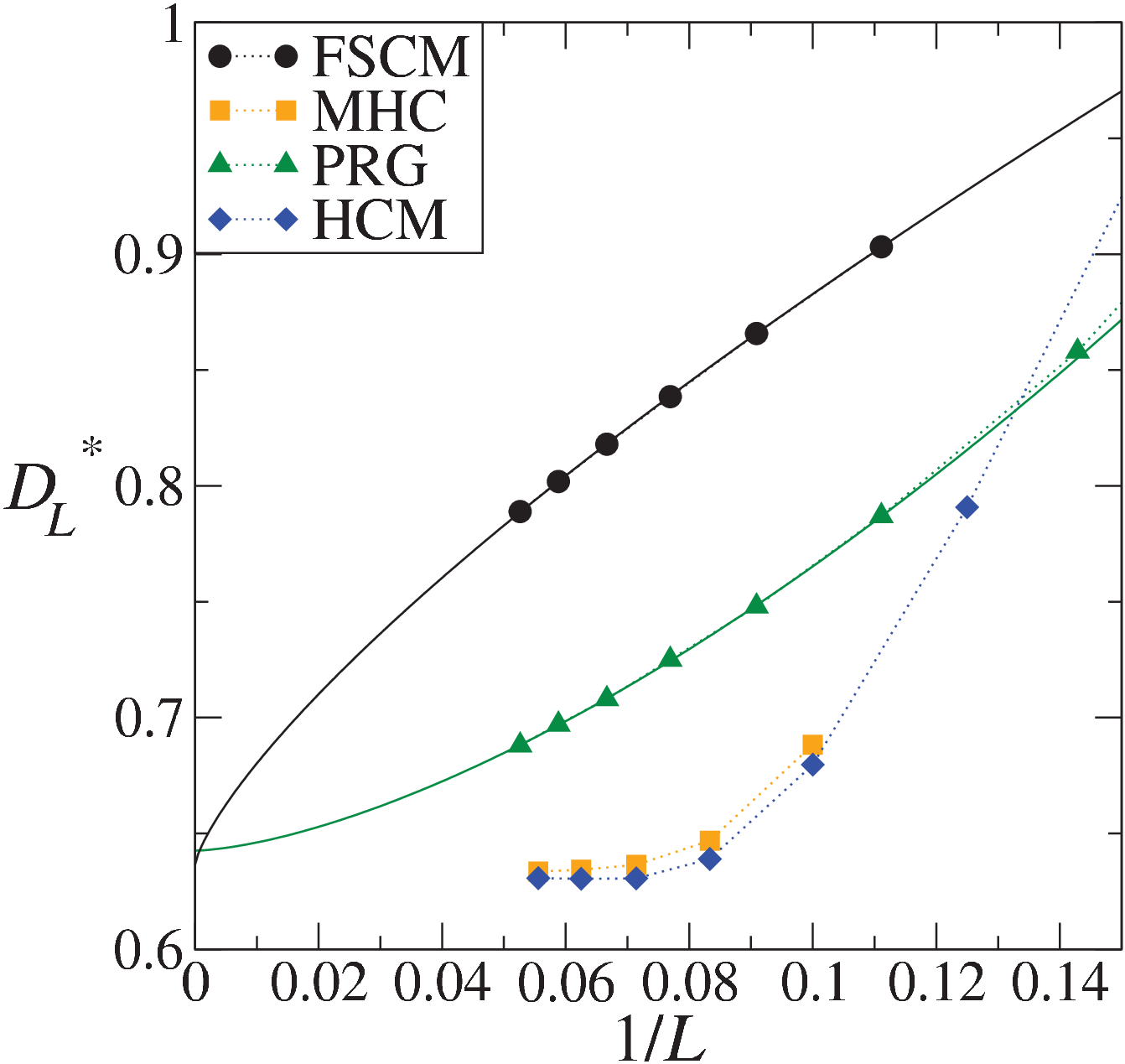}
\par\end{centering}

\caption{(color on-line) The sequences of pseudo-critical points obtained with
the FSCM, the PRG and new methods HCM (given by Eq.~(\ref{eq:HCML}))
and the MHC Eq.~(\ref{eq:MHC}). Continuous lines are algebraic best-fits
to the data.\label{fig:FSCM+PRG+hom}}

\end{figure}

The advantage of having fast converging sequences is that one can
limit the numerical calculations to short lattice sizes that can be
reached by exact diagonalization. To illustrate our procedures we
used exact Lanczos diagonalization for chains of length $L=8,10,12,14,16$
and used high precision DMRG simulations (with as many as $3^{7}$
optimized states and many zips) only for chain length $L=18,20$.
The role of $b_{L}$ is given here by $b_{L}=\langle\left(S_{i}^{z}\right)^{2}\rangle$.
We tested different methods: the FSCM, PRG, HCM, and MHC. We first
observed that the energy curves at different size do not cross. This
means that there are no linear terms in the scaling functions and
consequently we applied the FSCM and the HCM to the perturbation operator
$b_{L}$. The results are plotted in Figure \ref{fig:FSCM+PRG+hom}.
It is apparent that the HCM and MHC result in much more rapidly converging
sequences respect to other methods, including the PRG.

The extrapolated critical points together with the relative shift
exponents for different methods, are summarized in Table \ref{tab:Dc}.
In fact by inspecting the finite-size critical points we see that
for HCM and MHC the values of $D_{c}(L)$ at the largest plotted sizes
already are saturated within the error bar induced by the sampling,
$\delta D=0.1$, that propagates to $O(\delta D^{3})$ when the zeroes/crossings
are located using cubic splines (an error-free localization of zeroes would require adaptive sampling with increasing computational load). 
At this stage, if we insist on computing
a shift exponent, we can use a fit in which the infinite-size critical
point is kept fixed as in Table \ref{tab:Dc} and in this way we estimate
$\theta_{HCM},\,\theta_{MHC}$ $\ge8$. As far as FSCM and PRG are
concerned, the nonlinear best-fit curves underlying the data in Table
\ref{tab:Dc} are also displayed in Figure \ref{fig:FSCM+PRG+hom}.

\begin{table}
\begin{centering}
\begin{tabular}{|c|c|c|c|c|}
\hline 
 & FSCM  & PRG  & HCM  & MHC\tabularnewline
\hline 
\hline 
$D_{c}$  & $0.638\pm0.006$  & $0.643\pm0.003$  & $0.631\pm0.001$  & $0.634\pm0.001$\tabularnewline
\hline 
fitted $\theta$  & $0.75$  & $1.54$  & $\ge8$  & $\ge8$\tabularnewline
\hline 
\end{tabular}
\par\end{centering}

\caption{Extrapolated critical points together with shift exponents for various
methods.\label{tab:Dc}}
\end{table}

\subsection{BKT transition\label{sub:J1J2}}

To finish our series of numerical test we consider the example of a BKT transition. 
In such a transition the gap opens up so slowly (remind $\Delta\sim\xi^{-1}\sim\exp\left(-at^{-\sigma}\right)$)
that the PRG method is practically not applicable. As an example of
BKT transition we consider the spin-1/2 Heisenberg model with frustration
due to next-to-nearest neighbors interaction:
\[
H=\sum_{j=1}^{L}J_{1}\vec{S_{j}}\cdot\vec{S}_{j+1}+J_{2}\vec{S_{j}}\cdot\vec{S}_{j+2}.
\]
The model is equivalent to a 2-legs zigzag ladder with $L/2$ rungs.
We will also fix $J_{1}=1$ for clarity. The model is gapless for
$J_{2}<J_{2c}$ while it has a doubly degenerate ground state for
$J_{2}>J_{2c}$. The most precise estimate for the critical point was done
by Okamoto and Nomura in \cite{okamoto92} using a model-specific
method where the crossing points are determined by the coincidence of the gaps in the singlet and the triplet sectors. 
They found $J_{2c}=0.2411\pm0.0001$, which sets the ``accepted value'' to date. 
The role of the parameter $g$ is played here by $J_{2}$ while the perturbation operator is
given by $b_{L}=\langle\vec{S_{j}}\cdot\vec{S}_{j+2}\rangle$. The
numerical calculations were done using ladders with PBC and up to
$L/2=13$ rungs, an effective $\delta L=2$, and a number of DMRG states large enough to guarantee an exact value 
of the ground state energy up machine precision.
Here we have chosen to determine adaptively the mesh of values for $J_{2}$ in order to detect the zero-crossing points 
with the best precision. 

\begin{figure}
\noindent \begin{centering}
\includegraphics[height=6cm,clip]{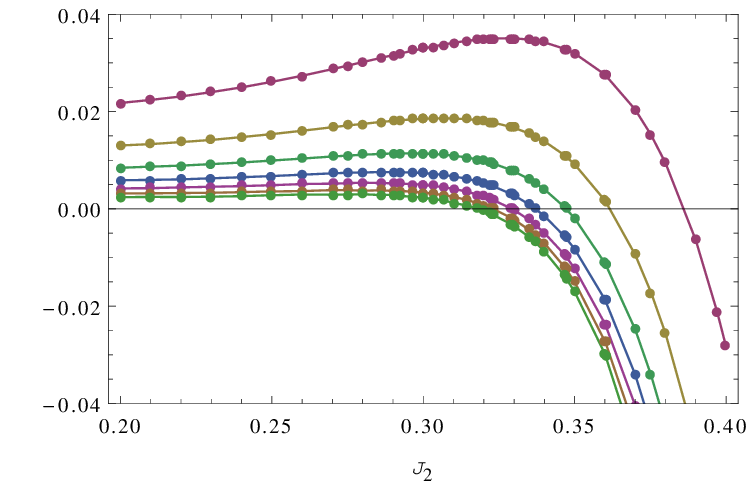}
\par\end{centering}
\noindent \begin{centering}
\includegraphics[height=8cm,clip]{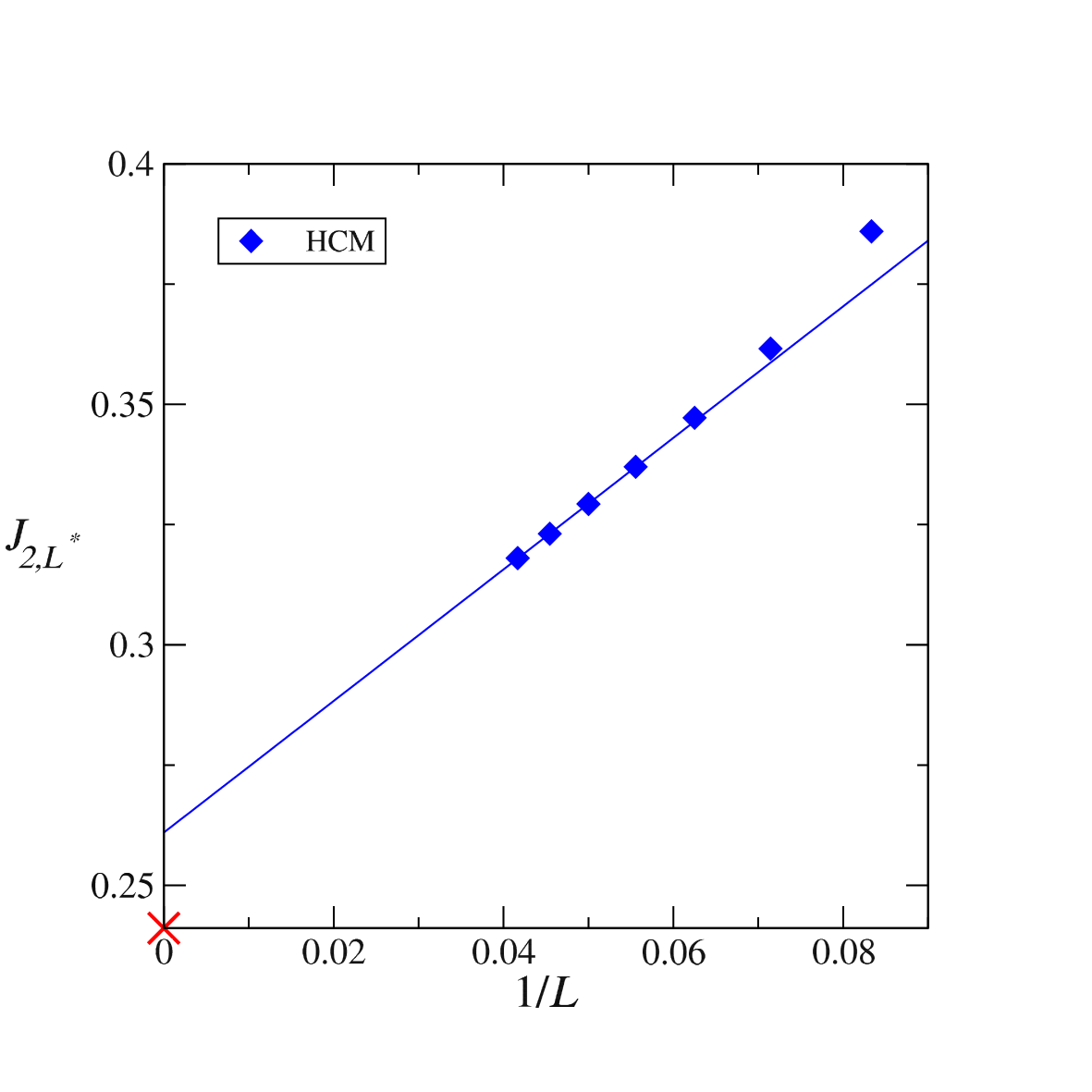}
\par\end{centering}

\caption{(color on-line) Upper panel: Search for the zeros of Eq.~(\ref{eq:HCML})
with $\delta L=2$. Lower Panel: Extrapolations of the pseudo-critical
points $J_{2,L}^{\ast}$. For reference, we have marked with a red cross the value $0.2411$ calculated in Ref. \cite{okamoto92} 
(see text for details).
\label{fig:homBKT}}

\end{figure}

We now pass to examine the different methods. As already mentioned
the PRG is not feasible because of the extremely slow opening of the
gap. The FSCM also fails as it turns out to be impossible to find
a real solution to the FSCM equation. Instead the HCM shows a
sequence of zero-crossing points, as displayed in the upper panel of Figure \ref{fig:homBKT}. 
This sequence, plotted in the lower panel of Figure \ref{fig:homBKT} as a function of $1/L$, 
converges to a finite $J_2$ with a shift exponent close to one.
A linear fit gives $J_{2c}=0.2610\pm0.0004$. Such a critical value is slightly larger than the one reported in
\cite{okamoto92}, but it is definitely different to it as the small error bars exclude compatibility between the two. 
According to the present numerical analysis, our method predicts a larger critical phase $J<J_2$, where the quantity $b(L)$ 
is found to be homogeneous of degree 2, a feature that is associated with a critical phase. 

In principle, both the HCM method and the one proposed by Okamoto and Nomura should converge to the same critical point, 
no matter how fast they attain the asymptotic value for $L\to \infty$. 
This discrepancy raises an interesting puzzling question on how to establish the right methods 
to determine the correct location of the BKT transition. 
Let us mention that the crossing method in \cite{okamoto92} is ruled by the additional assumption that in the RG scheme the 
bare coefficient of the marginal term vanishes at the BKT point. However, irrelevant terms at criticality may introduce a finite contribution to this coefficient, introducing a possible source of systematic deviation from the exact critical value.  
In any case, the solution of this challenging problem seems to need a thorough analysis close to a criticality, with a careful 
treatment of any possible logarithmic correction that is typically associated to transitions of BKT type.  
We think that this task deserves a dedicated study and it is clearly beyond the scopes of the present article. 
Here, we limit ourselves to report on the success of the HCM method to converge remarkably fast to a finite critical 
value for the BKT transition, a problem known to be extremely severe to treat.

\section{Conclusions}

Locating the critical point is one of the first issues which arise
when considering a statistical mechanical problem. 
Available methods to extract critical points from numerical simulations are,
among others: i) looking for the maxima of a susceptibility ii) intersection
points of reduced (Binder's) cumulants at different system sizes,
iii) look for closing of gaps in quantum systems (PRG). These methods
produce sequences of pseudo-critical points $g_{L}^{\ast}$ which
converge to the true critical point as $g_{L}^{\ast}=g_{c}+O\left(L^{-\theta}\right)$
with a leading algebraic term controlled by the shift exponent $\theta$.
The general belief is that the shift exponent is given by the inverse
of the correlation length exponent: $\theta=1/\nu$. In this article
we have developed accelerated methods for the location of critical points.
According to the present scaling analysis, our methods can in principle be applied at any second order transition 
in any dimension quantum or classical, temperature or parameter driven. The resulting
pseudo-critical sequences are characterized by a large value of the
shift exponent which enables these methods to be efficient even in
the extremely hard case of BKT transition where formally ''$\nu=\infty$''. 

The advantage of having such methods at disposal, is that one can
limit the analysis to much shorter system sizes. 
In this paper, we have shown how the proposed methods are able to produce fast converging sequences  
in 1D, focussing on three problems of different nature: the exactly solvable XY model, a non
integrable spin-$1$ model and the frustrated spin-$1/2$ Heisenberg model. 
We have shown that the value of the critical point is correctly determined in the first case, and in the second it is 
estimated with high accuracy. Interestingly, the Homogeneity Condition Method is able
to produce fast converging sequences in correspondence of the Berezinskii-Kosterlitz-Thouless
transition for the third test model, but the critical point value differs from the previous estimation given in Ref. \cite{okamoto92}. 
As both values are coming from scaling assumptions predicted by a renormalization group analysis, we take 
this discrepancy as a challenging problem to address in future studies. 
Finally, we think that our methods can find important applications in two dimensional quantum mechanical 
problems with frustration, for which quantum Monte Carlo methods are plagued with the sign problem
and the reliable numerical data are therefore restricted to very short sizes.

\ack{}{}

MR acknowledges support from Compagnia di san Paolo. 

\bibliographystyle{jphysicsB}
\addcontentsline{toc}{section}{\refname}\bibliography{ref}

\end{document}